\documentclass[12pt]{revtex4}

\usepackage{amssymb, amsmath, amsfonts, latexsym, verbatim, mathrsfs}

\usepackage{graphicx}

\begin{document}

\title{ON THE COMPLETENESS OF QUANTUM MECHANICS AND THE INTERPRETATION OF THE STATE VECTOR}
\thanks{This research is supported by the ARO MURI grant W911NF-11-1-0268.}

\author{GianCarlo Ghirardi}
\email{ghirardi@ictp.it}
\affiliation{Department of Physics, University of Trieste, and the Abdus Salam ICTP, Trieste (Italy)}
\author{Raffaele Romano}
\email{rromano@iastate.edu}
\affiliation{Department of Mathematics, Iowa State University, Ames, IA (USA) }


\begin{abstract}

\noindent Recently, it has been argued that quantum mechanics is complete, and that
quantum states vectors are necessarily in one-to-one correspondence with the
elements of reality, under the assumptions that quantum theory is correct and that measurement settings
can be freely chosen. In this work, we argue that the adopted form of the free choice assumption
is stronger than needed. In our perspective, there are hidden assumptions underlying
these results, which limit their range of validity. We support our argument by a model for the bipartite
two-level system, reproducing quantum mechanics, in which the free will assumption is respected,
and different quantum states can be connected to the same state of reality.

\end{abstract}


\keywords{Hidden variables, free will}

\maketitle


\section{INTRODUCTION}

Quantum mechanics is  one of the most successful theories ever developed. It has been elaborated to account for  the phenomena of the microscopic world and it has been experimentally verified to an impressive
degree of accuracy. It has required a radical change in the scientific conception of nature particularly in connection with its indeterminism and nonlocality.
However, from a conceptual point of view, the theory is not immune
of drawbacks. The crucial point stays in the fact that  it appears more as a set of operational prescriptions to predict outcomes of prospective measurement procedures
 than  a coherent description of reality. Its range of validity is not clear, it relies
on two different kinds of evolution depending on the rather vague notion of measurement.
As a consequence of these problematic aspects it is not clear what is the conceptual status of the quantum state vector $\psi$, the mathematical entity associated to the state of any physical system. After the famous incompleteness
argument by Einstein, Podolski and Rosen~\cite{einstein}, the question of whether $\psi$ represents a
state of reality or rather a state of knowledge, as suggested by its updating following
a measurement procedure, remains unanswered.

In order to address some of these questions, several {\it ontological models} of quantum
mechanics have been introduced. These are theories predictively equivalent to quantum mechanics, but
providing a possibly richer description of  microscopic reality in terms of the so-called {\it ontic
state}, i.e., the most accurate specification of the physical state of the system~\cite{harrigan}.
In these theories, the state vector $\psi$ might  embody only partial information on the ontic state. In particular, it is associated to  a distribution $\rho_{\psi} (\lambda)$ on the space of the ontic variables,
with $\rho_{\psi} (\lambda) \geqslant 0$ and
\begin{equation}\label{norm}
    \int \rho_{\psi} (\lambda) d \lambda = 1 \qquad {\rm for \, all} \, \psi.
\end{equation}
The ontic state $\lambda$ contains all the information about the elements of reality of the underlying theory, and, accordingly,  it provides
the complete description of the state of the system~\footnote{To clarify the problem of attributing elements of physical reality to a physical system it is useful to make reference, once more, to the EPR paper. In the case in which $\psi$ is an eigenstate of an observable $\Omega$ pertaining to the eigenvalue $\omega_{k}$, we knows that the standard quantum theory predicts that a measurement of $\Omega$ gives with certainty the outcome $\omega_{k}$, and, on the basis of this fact, EPR claim that the system possesses $\Omega = \omega_{k}$ as an element of physical reality. In such a case, with reference to the ontic state $\lambda$, we must impose that the value $\Omega(\lambda)$ of the observable $\Omega$ must be defined and must coincide with $\omega_{k}$. Obviously, it may very well happen (and to which extent this actually occurs depends on the ontic theory one has in mind), that for the given $\psi$ and $\lambda$, the knowledge of $\lambda$ allows the specification of further elements of reality besides those based on the EPR criterion. Typically, $\psi$ might be an eigenstate of the $z$ spin component of a particle, but knowledge of $\lambda$ might allow one to associate elements of physical reality also to the $x$ and/or $y$ spin components.}. In principle, $\lambda$ can be not fully accessible. Ignorance of its precise value is deemed to be the origin of various of the aforementioned
problematic aspects of quantum mechanics.

In the past literature, ontological models were usually referred to as {\it hidden variables theories},
meaning that additional variables, absent in the standard quantum formalism (and for this reason named ``hidden''),
were expected to supplement the information given by $\psi$,  providing in this way a complete description of the physical situation. From the beginning,
this program had supporters as well as opponents. J. von Neumann proved that deterministic hidden variables models were
unviable~\cite{vonneumann}, but later D. Bohm provided an explicit, fully deterministic, hidden variable model predictively equivalent to quantum
mechanics~\cite{bohm0,bohm}.  Later J.S. Bell appropriately stressed that  von Neumann's impossibility theorem
was circular, since it was based on logically unnecessarily strong assumptions~\cite{bell0}.
He also proved that any theory predictively equivalent to quantum mechanics must violate some
inequalities, named {\it Bell's inequalities} after him, which are  satisfied by any
local theory~\cite{bell00}.

Recently, a new argument supporting the completeness of quantum mechanics has been presented in~\cite{colbeck0},
based only on two assumptions: that quantum mechanics is correct ($QM$), and that a precise mathematical condition ($FR$)  - which the authors believe to express the request that measurement settings can be
freely chosen - be satisfied. According to~\cite{colbeck0}, any completion of quantum mechanics which consistently reproduces its outcomes,
necessarily comes at the expenses of the free will of the observers in choosing the measurement settings. Here,
we argue that, similarly to the case of von Neumann argument against hidden variables theories, also this result is
inconclusive due to the  unappropriate formalization of the assumption ($FW$) of free will, which we will  make precise in what follows. Specifically, we provide evidence that the physically meaningful request of $FW$
is not correctly expressed in mathematical terms in~\cite{colbeck0}, since the constraint $FR$, which stays at the basis of the authors' argument, is unnecessarily stronger than $FW$.
For related comments see~\cite{ghirardi0,ghirardi00,ghirardi}.

Our criticism has relevant consequences concerning the meaning of the quantum state vector $\psi$
in ontological models. First of all, let us recall that
the distributions $\rho_{\psi} (\lambda)$ might have overlapping supports for different $\psi$, a property which characterizes
the so-called {\it $\psi$-epistemic models}. Alternatively, when the supports are always disjoint we are dealing with {\it $\psi$-ontic models}~\cite{harrigan}.
According to this classification the state vector $\psi$ acquires a strongly different conceptual status.
In fact, in a $\psi$-ontic theory different state vectors necessarily correspond to different ontic
states, whereas in a $\psi$-epistemic theory they could correspond to the same ontic state.
As a consequence, one can associate to $\psi$  well defined elements of reality only in the first case.

As it should be obvious, the distinction between $\psi$-epistemic and $\psi$-ontic models implies completely different
interpretations of the quantum state vector. For this reason, these classes of models have recently been the subject
of several investigations. A first argument arguing that $\psi$-epistemic models necessarily contrast
with quantum mechanics was developed in~\cite{pusey}, but its conclusions are  severely limited by the assumption that factorized
quantum states correspond to factorized states of the underlying theory. In fact, an explicit $\psi$-epistemic
model has been built by relaxing this assumption~\cite{lewis}. Another no-go theorem for $\psi$-epistemic models has been developed
in~\cite{colbeck}, based on the completeness result presented in~\cite{colbeck0}. Our arguments against the assumption  $FR$
make also this argument inconclusive in determining the ontological status of the quantum state vector,
since a weaker and more significant condition of free will is still compatible with $\psi$-epistemic models. We substantiate our
argument by presenting an explicit $\psi$-epistemic model for a pair of two-level systems, in which
the experimenters can freely choose their local settings, so that $FW$ holds, even though condition $FR$ is violated.

Before proceeding, a relevant remark is at order. Here, and anywhere in this work, when using the
term {\it reality} we do not mean {\it what the world is}, but rather {\it what the theory is about},
exactly in the  spirit of Bell's beables~\cite{bell}:
\begin{quote}
{\it I use the term 'beable' rather than some
more committed term like 'being' or 'beer'
to recall the essentially tentative nature of
any physical theory.}
\end{quote}
We further assume that a complete description can be obtained by confining our attention to
a specific set of physical systems, the rest of the universe not contributing at all,
as appropriately required by Shimony, Horne and Clauser~\cite{shimony}:
\begin{quote}
{\it Unless we proceed under the assumption
that hidden conspiracies [...]
do not occur, we have abandoned in advance
the whole enterprise of discovering the laws
of nature by experimentation.}
\end{quote}
Accordingly, we do not consider superdeterministic models in our analysis.

In Section~\ref{sec2} we review the completeness argument of~\cite{colbeck0}, and its
implications on the interpretation of the state vector~\cite{colbeck}, by slightly
modifying the original notations for sake of clarity. We only describe the steps which are relevant
to our analysis, and refer the interested reader to the original papers for further details. In Section~\ref{sec3} we
criticize the assumption that the condition  $FR$ introduced in~\cite{colbeck0} correctly embodies the request $FW$ of free will. Actually, we shall prove that $FR$ implies more than
the free will of the experimenters, and this will make clear how our analysis limits the results of~\cite{colbeck0,colbeck}.
In Section~\ref{sec4} we substantiate our arguments by providing an explicit model in which the quantum state
vector is not in one-to-one correspondence with the elements of reality. Finally, in Section~\ref{sec5},
we criticize a more recent argument supporting the completeness of quantum mechanics, and we
present our final remarks.

\section{THE COMPLETENESS ARGUMENT OF COLBECK AND RENNER}\label{sec2}

We consider the standard EPR-like scenario: two space-like separated observers
perform local measurements on the two parties of an entangled state $\psi$. The
measurement settings on the two sides are given by vectors $A$ and $B$, and the measurement outcomes
by $X$ and $Y$, whose values are in $\{-1, 1\}$. We assume that additional information on the ontic
state $\lambda$ can be accessed through a measurement procedure with setting $C$ and
output $Z$. All these quantities are assumed to be space-time random variables, whose values can be
associated to well-defined space-time points $(t, r_{1}, r_{2}, r_{3})$, making precise the statement that two
measurement procedures and the associated outcomes are space-like with respect to one another. All these
quantities are described by joint conditional probabilities of the form $P_{XYZ|ABC}$, and similar ones.
Generally, these probabilities are conditioned by $\psi$, but usually this dependence will be left implicit
for notational convenience.

The argument presented in~\cite{colbeck0} relies essentially on two assumptions, which we now review. The first one,
denoted by $FR$, is that the observers can independently choose the measurements they perform. As explicitly stated in~\cite{colbeck0}:
\begin{quote}
{\it Assumption $FR$ is that the input, $A$, of a measurement process can be chosen such that it is uncorrelated with certain other
space-time random variables, namely all those whose coordinates lie outside the future light-cone of the coordinates of $A$.}
\end{quote}
This assumption holds true for $B$ and $C$ as well, and it is mathematically expressed (and this is actually what the author's mean by the request $FR$) through
the following requirements on the conditional probabilities:
\begin{equation}\label{FR}
P_{A|BCYZ} = P_A, \qquad
P_{B|ACXZ} = P_B, \qquad
P_{C|ABXY} = P_C.
\end{equation}
The second assumption, denoted by $QM$, is that  quantum mechanics is correct~\cite{colbeck0}:
\begin{quote}
{\it Measurement outcomes obey quantum statistics and [...] all processes within quantum theory can be considered as
unitary evolutions, if one takes into account the environment. [Moreover] the second part of the assumption
need only hold for microscopic processes on short timescales.}
\end{quote}
Now, by  relying only on these two assumptions, the authors of~\cite{colbeck0} prove that any knowledge
on the ontic state besides the one given by $\psi$ is useless, since it cannot modify the outcome distributions
predicted by quantum mechanics: $P_{XY|ABCZ} = P_{XY|AB}$, $P_{X|ABCZ} = P_{X|A}$ and  $P_{Y|ABCZ} = P_{Y|B}$. They conclude that~\cite{colbeck0}
\begin{quote}
{\it Any attempt to better explain the outcomes of quantum measurements is destined to fail. Not only is the
universe not deterministic, but quantum theory provides the ultimate bound on how unpredictable it is.
[...] under the assumption that measurement settings can be chosen freely, quantum theory really is complete.}
\end{quote}
The technical details
of their proof are not relevant here, since our criticism focuses on the adoption of the
assumption $FR$ to formalize the request $FW$ of the possibility of freely choosing the apparatus settings. We point out that all the three constraints in (\ref{FR}) are needed to prove
this result and that  these constraints imply the non-signalling conditions:
\begin{equation}\label{NS}
    P_{YZ|ABC} = P_{YZ|BC}, \qquad P_{XZ|ABC} = P_{XZ|AC}, \qquad P_{XY|ABC} = P_{XY|AB},
\end{equation}
so that no experimenter can extract, form his outcomes, any information about the measurement
settings which have been chosen by the other experimenters. This fact is physically relevant since it
avoids  superluminal communication between them.

By adapting their argument, in~\cite{colbeck} the same authors provide a no-go theorem for $\psi$-epistemic
models, ultimately based on the same assumptions $FR$ and $QM$. They show that $P_{X|\lambda A} = P_{X|\psi A}$,
and then conclude that there should be a one-to-one correspondence between $\lambda$ and $\psi$. As before, the
technical machinery leading to this conclusion is irrelevant here, but we stress once more that also this result
is a direct consequence only of the two assumptions: $FR$ and $QM$. In the following, we assume that $QM$ is fulfilled,
and we criticize the identification of $FR$ with the assumption $FW$ of the free choice of measurement settings, arguing that
the latter can be satisfied even though $FR$ is violated.

\section{CRITICISM OF THE FR ASSUMPTION AND ITS CONSEQUENCES}\label{sec3}

As noticed by the authors of~\cite{colbeck0}, their result has meaningful consequences when applied to
existing hidden variables models, notably, to Bohmian mechanics. Since this is a deterministic completion
of quantum mechanics, following the analysis of J.P. Jarrett, its non-locality necessarily arises from a violation
of the condition of parameter independence ($PI$), which is expressed as
\begin{equation}\label{PI}
    P_{X|AB\lambda} = P_{X|A\lambda}, \qquad P_{Y|AB\lambda} = P_{Y|B\lambda}.
\end{equation}
In the context of~\cite{colbeck0}, $PI$ reduces to $P_{X|ABZ} = P_{X|AZ}$ and $P_{Y|ABZ} = P_{Y|BZ}$,
since it is assumed that the information represented by $\lambda$ is fully accessible (and obtained by $Z$). This is
by itself a constraint which limits the applicability of the results presented in~\cite{colbeck0},
since it ignores the important  distinction between controllable and uncontrollable hidden variables, which has been strongly emphasized by A. Shimony~\cite{shimony2}.
In fact, it is generally assumed that the hidden variables are not completely accessible;
that is, there can be some physical principles which limit their knowledge and their manipulation, ensuring, first of all,
the impossibility of superluminal signalling. Maintaining the authors' perspective on the accessibility of $\lambda$,
we observe that the non-signalling conditions (\ref{NS}) automatically imply $PI$, and therefore the assumption $FR$ trivially forbids all the deterministic
completions of quantum mechanics. In~\cite{colbeck0}, this fact is interpreted by
claiming that these models violate the assumption of free choice. Nonetheless, as we have already mentioned and as we shall shortly prove, the $FR$
assumption cannot be considered as the appropriate formalization  of the free will assumption $FW$, since it embodies more than
this assumption. Therefore, the violation of $FR$ cannot be automatically related to lack of free will.

To start with, we introduce the condition $FW$, which,
in our opinion, provides a more suitable expression for the free choice assumption. We observe that, when dealing with correlation experiments of this type, one is merely interested in the freedom of choice of the measurement settings $A$ and $B$, located at the two wings of the experimental set-up. This condition, when
expressed in terms of conditional probabilities, reads
\begin{equation}\label{FR'}
    P_{A|B\lambda} = P_A, \qquad P_{B|A\lambda} = P_B,
\end{equation}
and it fully accounts for the fact that the two experimenters can freely and independently choose which observable to measure,
since it implies the factorization $P_{AB\lambda} = P_A P_B P_{\lambda}$. Notice that the assumption $FW$ does not involve a third party, and its measurement setting $C$.
Actually, the authors of~\cite{colbeck0} remark that the additional information supplementing $\psi$
\begin{quote}
{\it [...] must be static, that is, its behavior cannot depend on where or when it is observed [...] so, we can consider
the case where its observation is also space-like separated from the measurements specified by $A$ and $B$.}
\end{quote}
This statement intends to stress that $FR$ is the appropriate way to express the free will request, since it is consistent with the condition $P_{C|ABXY} = P_C$.
Nonetheless, for future reference, we take  it as an independent assumption, denoted as $ST$,
and expressed through the condition
\begin{equation}\label{ST}
    P_{CZ|ABXY} = P_{CZ}.
\end{equation}
Notice that in~\cite{colbeck0} $ST$ does not appear as an assumption, nor the mathematical expression (\ref{ST}) appears somewhere: it is
introduced by us in our criticism of the assumption $FR$.

There is another striking difference between the conditions $FR$ and $FW$: the former involves the random variables
$X$ and $Y$, the latter does not. We believe that it is important to avoid this dependence, since
the extra variables $X$ and $Y$ could bring spurious correlations, completely independent of the free will assumption.
To clarify the idea, let us assume that we are interested in the free choice of $A$, and let us raise the following question: if $P_{A|BCYZ} \ne P_A$,
can we conclude that $A$ cannot be freely chosen? We do not think that this is the case. For
instance, we might suppose that the two Stern-Gerlach apparatuses located at the opposite
wings of the experiment could superluminally communicate, at a suitable finite speed. In
the rest frame of the two experimenters, the communication happens after the free choice
of $A$ and $B$, and before the generation of the outputs $X$ and $Y$ . In this way, correlation
between $A$ and $Y$ might be produced. Even though it is highly implausible that a physical
process as the one just mentioned has any physical meaning, its consideration has nothing
to do with the fact that $A$ and $B$ can be freely and independently chosen. In other words, we believe that a good mathematical formulation of the free measurements choice should
not automatically reject situations where free will and superluminal communication coexist. Unfortunately,
this is not the case when the $FR$ assumption is considered, since it implies the non-signalling constraints (\ref{NS}),
as stated in Section~\ref{sec2}.

In accordance with the previous analysis, we find convenient to write non-signalling conditions (which are independent from the additional
information on the ontic state, and from the setting of its measuring device) as follows:
\begin{equation}\label{NS2}
    P_{X|AB} = P_{X|A}, \qquad P_{Y|AB} = P_{Y|B}.
\end{equation}
They are weaker than relations (\ref{NS}), still they fully express the impossibility
of superluminal communication in the standard EPR scenario. In the following, we refer to Eq. (\ref{NS2})
as the $NS$ assumption. As proven in~\cite{ghirardi}, the relation between $FR$, $FW$, $ST$ and $NS$ is given by
\begin{equation}\label{rel}
    FW \wedge NS \wedge ST \Rightarrow FR,
\end{equation}
and, if the ontic state $\lambda$ is fully accessible, also the inverse implication holds,
\begin{equation}\label{rel2}
     FR \Rightarrow FW \wedge NS \wedge ST.
\end{equation}
Therefore, in the case considered in~\cite{colbeck0}, the $FR$ assumption is equivalent to the logical
conjunction of our assumptions $FW$ of free will, staticity $ST$ of the ontic state, and impossibility of superluminal
communication, $NS$. This result proves our statement that $FR$ is more than the free choice assumption, and
shows that, even if the ontic state is (at least partially) unaccessible, negation of $FR$ does
not necessarily imply absence of free will. It might depend on a violation of free will, on the fact that
the additional information on $\lambda$ is not static, or on a violation of the impossibility of superluminal communication. Of those, the second
condition appears the easiest to digest: why the extra information on $\lambda$, which is complementing
the information provided by the state vector $\psi$, should be static, when $\psi$ itself, and the
measurement procedure involved in its preparation, are not space-like separated with respect to $A$
and $B$?

Therefore, the completeness argument presented in~\cite{colbeck0} has not the claimed generality,
although it is formally correct. When considering Bohmian mechanics, the
statement that measurement settings cannot be freely chosen is not justified. If we
assume that $\lambda$ is fully accessible (that is, that all positions
are known), it is a well known fact that the theory allows superluminal communication.
Otherwise, it is likely that the additional information on the ontic state is not static
(in fact, it should be a partial information on the positions, which are distributed according to
$\psi$, which in turn is certainly a non-static quantity). In both cases, violation of $FR$ does not
require lack of free will.

The no-go theorem for $\psi$-epistemic theories derived in~\cite{colbeck} is based on the
assumptions $QM$ and $FR$. Therefore, the same conclusions apply to this case as well,
and this result is not general at all. Both the questions of whether quantum mechanics is a
complete theory, and the interpretation of the quantum state as a state of knowledge or an element
of reality, remain open. In the following section we present a $\psi$-epistemic model for a pair of
two-level systems, fully consistent with quantum mechanics, in which the measurement settings $A$
and $B$ can be freely chosen.

\section{A $\psi$-EPISTEMIC MODEL FOR THE BIPARTITE SYSTEM}\label{sec4}

The model is a trivial application of the results of~\cite{lewis}, where, however,
it has not been realized that the family of $\psi$-epistemic models therein presented
are valid beyond the single system case. Following our reasoning in the previous section,
we can claim that superluminal influences of measurement choices upon ontic variables,
in the multipartite case, are not necessarily expected. Despite in~\cite{lewis} a
system of arbitrary dimension is considered, here we limit our attention to a pair of
two-level systems for sake of simplicity.

The model is inspired by a generalization, to the case of two subsystems, of a Bell model
for a single two-level system. We start by describing this generalization.
If we denote by $\mathcal{H}$ the Hilbert space of the single
system, an arbitrary quantum state vector is given by $\varphi \in \mathcal{H} = \mathcal{H}_A \otimes \mathcal{H}_B$.
The hidden variable is a real parameter $\tau \in T = [0,1]$, and the ontic state is given by
the ordered pair $\lambda = (\varphi, \tau) \in \mathcal{H} \times T$. Its distribution
corresponding to the quantum state $\psi$ is given by
\begin{equation}\label{Dont}
    \rho_{\psi} (\lambda) = \rho_{\psi} (\varphi, \tau) = \delta (\varphi - \psi),
\end{equation}
that is, $\tau$ is uniformly distributed over $T$ for all states $\psi$. We consider now the local
observables $\hat{A} = A \cdot \sigma_A$ and $\hat{B} = B \cdot \sigma_B$, where the settings $A$ and $B$
are unit real vectors in $\mathbb{R}^3$ and $\sigma_A$, $\sigma_B$ are the vectors of Pauli matrices
acting on $\mathcal{H}_A$ and $\mathcal{H}_B$ respectively. We denote the common eigenvectors
of the commuting observables $\hat{A}$, $\hat{B}$ and $\hat{A} \otimes \hat{B}$ as $\phi_j$, and we order them
according to
\begin{equation}\label{ord}
    \vert \langle \phi_j \vert 0 \rangle \vert^2 \geqslant \vert \langle \phi_{j + 1} \vert 0 \rangle \vert^2 \qquad {\rm with} \;\; j = 0, \ldots, 3,
\end{equation}
where $\vert 0 \rangle \in \mathcal{H}$ is an arbitrary reference state. The states $\phi_j$ are
factorized, and they are associated with well defined outcomes $(X, Y)$ for the observables $\hat{A}$
and $\hat{B}$, which are in the set $\{ (-1,-1), (-1,1), (1,-1), (1,1) \}$. The outcome of $\hat{A} \otimes
\hat{B}$ is given by the product $XY$. The specific correspondence between $j$ and the pairs of possible
outcomes for $X$ and $Y$ depends on the local observables taken into account, in the way we describe now.
In the ontic state $\lambda = (\varphi, \tau)$ the value of the projectors $\hat{\Phi}_j = \vert \phi_j \rangle \langle \phi_j \vert$ are defined as
\begin{equation}\label{assign}
    \Phi_j (\lambda) = 1, \qquad {\rm if} \;\; \sum_{k = 0}^{j - 1} \vert \langle \phi_k \vert \varphi \rangle \vert^2 \leqslant \tau < \sum_{k = 0}^{j} \vert \langle \phi_k \vert \varphi \rangle \vert^2,
\end{equation}
and $\Phi_j (\lambda) = 0$ otherwise; when j = 0, the lower bound for $\tau$ is defined to be 0
(in this case the former expression is meaningless).
Therefore, with any $\lambda$ there is associated a value of $j$. Since the spectral decomposition of the local operators $\hat{A}$ and $\hat{B}$ (as well as that of $\hat{A} \otimes \hat{B}$) can be written through the projectors $\hat{Phi}_j$, the corresponding outcomes $X = X(\lambda)$ and $Y = Y(\lambda)$ (and, of course, their product) are unambiguously determined. They are distributed according to the quantum mechanical rules~\cite{lewis}, since
\begin{equation}\label{consistency}
    \int d \lambda \rho_{\psi} (\lambda) \Phi_j (\lambda) =
    \int d \tau \Phi_j (\psi, \tau) = \vert \langle \phi_j \vert \psi \rangle \vert^2.
\end{equation}

As the original Bell model for a single two-level system, this generalized model is $\psi$-ontic. In fact,
according to (\ref{Dont}), the distributions corresponding to different state vectors never overlap. However,
this property is not needed at all in order to produce a model predictively equivalent to quantum mechanics:
it is sufficient that the measure of the support of the function $\Phi_j (\lambda)$ over $\{ (\varphi, \tau), \tau \in T \}$ equals the quantum probability $\vert \langle \phi_j \vert \psi \rangle \vert^2$. Following~\cite{lewis}, by suitably redistributing this support, it is possible to turn the model into a
$\psi$-epistemic one. After noticing that $\vert \langle \phi_0 \vert 0 \rangle \vert^2 \geqslant 1/4$,
the authors introduce
\begin{equation}\label{z}
    z(\varphi) = \inf_{\phi: \langle \phi \vert 0 \rangle \vert^2 \geqslant 1/4} \vert \langle \phi \vert \varphi \rangle \vert^2,
\end{equation}
which is used to define a subset of the ontic space,
\begin{equation}\label{subset}
    \mathcal{E}_0 = \{(\varphi, \tau): \vert \langle \varphi \vert 0 \rangle \vert^2 > 3/4 \;\; {\rm and} \;\; 0 \leqslant \tau < z(\varphi) \}.
\end{equation}
This set has the property that all $\lambda \in \mathcal{E}_0$ are in the support of $\Phi_0 = \vert \phi_0 \rangle \langle \phi_0 \vert$. Therefore, these ontic states have all the same outcome, which is the one associated with $\phi_0$. If we redistribute over the whole of $\mathcal{E}_0$ the probabilities which the distribution $\rho_{\psi} (\lambda)$ in (\ref{Dont}) assigns to ontic states within $\mathcal{E}_0$ itself,
we achieve the desired target. For example, by denoting by $\rho_{\mathcal{E}_0} (\lambda)$ the uniform distribution over $\mathcal{E}_0$, and by $\Theta(x)$ the unit-step function, such that $\Theta(x) = 1$ for $x \geqslant 0$ and $\Theta(x) = 0$ for $x < 0$, it is possible to define
\begin{equation}\label{Depi}
    \tilde{\rho}_{\psi} (\lambda) = \tilde{\rho}_{\psi} (\varphi, \tau) =
    \left\{
      \begin{array}{ll}
        \delta (\varphi - \psi), & \hbox{\quad for $\vert \langle \psi \vert 0 \rangle \vert^2 \leqslant 3/4$} \\
        \delta(\varphi - \psi) \Theta(x - z (\psi)) + z(\psi) \rho_{\mathcal{E}_0} (\lambda), & \hbox{\quad for $\vert \langle \psi \vert 0 \rangle \vert^2 > 3/4$}
      \end{array}
    \right.
\end{equation}
which, by construction, satisfies
\begin{equation}\label{consistency2}
    \int d \lambda \tilde{\rho}_{\psi} (\lambda) \Phi_j (\lambda) =
    \vert \langle \phi_j \vert \psi \rangle \vert^2,
\end{equation}
but in general produces overlapping supports for distinct state vectors. Therefore, if we replace
$\rho_{\psi} (\lambda)$ by $\tilde{\rho}_{\psi} (\lambda)$, we obtain a $\psi$-epistemic model for
a pair of two-level systems. There is not reason why the conditions $FW$ and $NS$ introduced in the
previous section should be violated, therefore we conclude that it is possible to build $\psi$-epistemic
models which do not ask for violations of the free will assumption, or superluminal communication.

In this section we have adopted a weak notion of $\psi$-epistemic model: there must be at least one pair of quantum state vectors whose corresponding distributions have overlapping supports. A stronger definition requires that all non-orthogonal quantum states have overlapping distributions. While a detailed treatment
of this case is out of the scopes of this paper, we point out that our remarks on the
expression of the free will assumption holds true also in this case.

\section{FINAL REMARKS AND CONCLUSIONS}\label{sec5}

Recently, the authors of~\cite{colbeck0,colbeck} have provided a different argument
leading to the same conclusion of completeness of quantum mechanics~\cite{colbeck2}. This work is
presented as an expanded and more pedagogical version of~\cite{colbeck0,colbeck}, nonetheless
it appears as a distinct derivation. For instance, the result that quantum mechanics is maximally
informative (that is, complete), asks for different resources in the cases of maximally and non-maximally
entangled states, unlike~\cite{colbeck0,colbeck}. Here we are not interested in a general
analysis of this work, but in investigating the assumption of free choice which appears in it, which
is the most significative requirement to draw the conclusions expressed in the paper.
Following the authors~\cite{colbeck2},
\begin{quote}
{\it [...] a parameter of the theory, say $A$, is considered free if it is possible
to choose $A$ such that it is uncorrelated with all other values (described by the theory)
except those that lie in the causal future of A.}
\end{quote}
They consistently express this idea through a factorization of the probabilities
$P_{A \Gamma_A} = P_A P_{\Gamma_A}$, where $\Gamma_A$ is the set of all random variables
which are not in the causal future of $A$. This condition must hold also for $B$,
that is $P_{B \Gamma_B} = P_B P_{\Gamma_B}$, where $\Gamma_B$ is the set of all random
variables which are not in the causal future of $B$.

Differently from the former assumption $FR$,
this condition does not involve the random variables $C$ and $Z$, and therefore it is
free from the former assumption of static variables, $ST$. Nonetheless, as $FR$ did,
also this condition automatically satisfies $NS$, and therefore it cannot coexist, neither
in principle, with a theory allowing superluminal communication. This can be derived
by expressing $P_{ABX}$ in different ways:
\begin{equation}\label{inc1}
    P_{ABX} = P_{X|AB} P_{AB} = P_{X|AB} P_A P_B,
\end{equation}
where, in the second step, we have considered that $B \in \Gamma_A$; but also
\begin{equation}\label{inc2}
    P_{ABX} = P_{AX} P_{B} = P_{X|A} P_A P_B,
\end{equation}
where the first step follows from $A \in \Gamma_B$ and $X \in \Gamma_B$. By comparing
(\ref{inc1}) and (\ref{inc2}), we derive the non-signalling condition $P_{X|AB} = P_{X|A}$;
a similar argument proves that $P_{Y|AB} = P_{Y|B}$, and finally $NS$ is derived.

Therefore, according to our former discussion, we believe that this is not a legitimate
expression of the free will, and also the results of this work are not as general as stated.
Notice that, to motivate their mathematical expression of the free will, the authors of~\cite{colbeck2}
mention how J.S. Bell affirmed characterized the concept of free variables~\cite{bell2}:
\begin{quote}
{\it for me this means that the values of such variables have implications
only in their future light cones.}
\end{quote}
However, ``having no implications'' is not the same as ``being uncorrelated''. The fact that I
decide to go out with my umbrella is correlated with the weather forecasts, but it has not implications
on them. Therefore, in our opinion, the expression of the free will adopted in~\cite{colbeck2} does not
expresses the Bell's point of view, and is not adequate.

For maximally entangled states, there is independent evidence that no theory consistent
with quantum mechanics can produce probabilities differing from those predicted by the
quantum theory~\cite{colbeck3,branciard,ghirardi2,dilorenzo}. Nonetheless, this result is in general false for non-maximally entangled states~\cite{ghirardi00}.
Notice that, in~\cite{colbeck2}, the predictive equivalence between the quantum theory and
the underlying model, in the case of non-maximally entangled states, is obtained by
reducing this case to that of maximally entangled states. This step relies on
the extra resource of {\it embezzling states}, a special set of states whose precise
characterization requires an infinite-dimensional Hilbert space~\cite{vandam}. By using these states, the
two experimenters can transform an arbitrary state of some Hilbert space into a maximally
entangled state of the same Hilbert space, via local operations. We believe that the
use of this resource is not legitimate in this context, since it does not allow a self-contained
description of any finite-dimensional Hilbert space.

We conclude that the issue of completeness of quantum mechanics, and the problem
of the interpretation of the quantum state vector, are still open.


\end{document}